%% file: mainv4.tex
\documentclass[aps,twocolumn,pra,superscriptaddress]{revtex4-2}

\usepackage{amsmath}

\usepackage{amssymb}

\usepackage{booktabs}
\usepackage{longtable}
\usepackage{array}

\newcolumntype{P}[1]{>{\raggedright\arraybackslash}p{#1}}

\usepackage{mathrsfs}
\usepackage{bm, dsfont}
\usepackage[usenames,dvipsnames]{color}
\usepackage{colortbl}
\usepackage[table]{xcolor}
\usepackage{enumitem}

\usepackage{graphicx}
\usepackage{natbib}
\usepackage[colorlinks=true,linkcolor=blue,citecolor=blue,urlcolor=blue]{hyperref}
\usepackage{cleveref}
\usepackage{hypcap}
\usepackage{verbatim, float}
\usepackage{psfrag}
\usepackage{tikz}
\usetikzlibrary{arrows.meta, automata,
                positioning,
                quotes,calc}
\usepackage[normalem]{ulem}
\usepackage{physics}		
\usepackage{amsmath}
\usepackage{amsthm}
\newcommand{\ii}{\mathrm{i}}
\newtheorem{conjecture}{Conjecture}

\newtheorem{theorem}{Theorem}
\newtheorem{observation}{Observation}

\usepackage{tikz}

\newcommand{\BlochTet}{%
  \mathchoice
    {\tikz[baseline=-0.6ex,scale=0.20]{\BlochTetTikz}}
    {\tikz[baseline=-0.4ex,scale=0.18]{\BlochTetTikz}}
    {\tikz[baseline=-0.3ex,scale=0.14]{\BlochTetTikz}}
    {\tikz[baseline=-0.22ex,scale=0.11]{\BlochTetTikz}}%
}

\newcommand{\BlochTetTikz}{%
  \draw[thick] (0,0) circle(1.3);
  \coordinate (A) at (0,0);             
  \coordinate (B) at (0,1);             
  \coordinate (C) at (-0.866,-0.5);     
  \coordinate (D) at (0.866,-0.5);      
  \draw[thin] (A)--(B) (A)--(C) (A)--(D);
  \draw[thin] (B)--(C) (C)--(D) (D)--(B);
  \foreach \p in {A,B,C,D}{
    \fill (\p) circle(0.1);
  }
}
\begin{document}
\author{Jef Pauwels}
\author{Nicolas Gisin}
\affiliation{Department of Applied Physics, University of Geneva, Switzerland}
\affiliation{Constructor University, 28759 Bremen, Germany}

\title{The Multiqubit Elegant Joint Measurement}

\begin{abstract}
The Elegant Joint Measurement (EJM) is a highly symmetric, partially entangled two-qubit measurement whose local marginals form a regular tetrahedron on the Bloch sphere and which has a low entanglement cost for local implementation. It plays a central role in quantum networks exhibiting nonclassical correlations and serves as a paradigmatic example of an entangled measurement with local structure. Despite its significance, generalizing the EJM beyond two qubits has remained unresolved. Here, we extend the EJM to the multipartite setting by identifying all tetrahedrally symmetric, efficiently localizable multiqubit bases. For two qubits, these criteria uniquely select the EJM. For three or more, they yield a discrete set of equivalence classes, reflecting the richer structure of multiparticle entanglement.
\end{abstract}
\maketitle

\section{Introduction}

While entangled states have been studied extensively, the structure of entangled measurements--and their role in quantum protocols--remains comparatively under-explored~\cite{Cavalcanti2023}. The Elegant Joint Measurement (EJM) provides a concrete and highly symmetric example of such a measurement, where entanglement coexists with strong local structure.

\begin{figure}[t]
\centering
\includegraphics[width=0.20\textwidth]{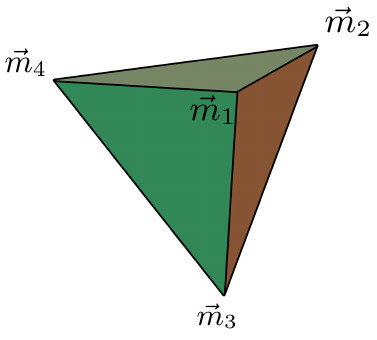}
\caption{\emph{Local Bloch structure of the Elegant Joint Measurement.}
The local Bloch vectors of the EJM basis point to the vertices of a regular tetrahedron:
$\vec{m}_1 = \frac{1}{\sqrt{3}}(1,1,1)$, 
$\vec{m}_2 = \frac{1}{\sqrt{3}}(1,-1,-1)$, 
$\vec{m}_3 = \frac{1}{\sqrt{3}}(-1,1,-1)$, and 
$\vec{m}_4 = \frac{1}{\sqrt{3}}(-1,-1,1)$.}
\label{fig:tetrahedron}
\end{figure}

The EJM was first formally introduced in Ref.~\cite{Gisin2019}, though it appeared in earlier forms in different contexts \cite{Gisin1999}. It is defined as the two-qubit orthonormal basis
\begin{equation} \label{eq:EJM}
\ket{\psi_{\mathrm{EJM},i}}= \frac{\sqrt{3}+1}{2\sqrt{2}}\ket{\vec{m}_i,-\vec{m}_i}+\frac{\sqrt{3}-1}{2\sqrt{2}}\ket{-\vec{m}_i,\vec{m}_i}\,,
\end{equation}
where \( \ket{\vec{m}_i} \) and \( \ket{-\vec{m}_i} \) denote the \( \pm 1 \) eigenstates of the qubit observable \( \vec{m}_i \cdot \vec{\sigma} \), whose Bloch vectors \( \vec{m}_i \) form the vertices of a regular tetrahedron (see Fig.~\ref{fig:tetrahedron}).

The EJM has several appealing properties. It is iso-entangled, meaning that all basis states share the same degree of entanglement. It is only partially entangled, so it has both nonlocal and local structure. Remarkably, this local structure is, in a sense, maximal: the qubit marginals point to the vertices of a regular tetrahedron on the Bloch sphere, i.e., they form the four directions associated with a qubit SIC geometry~\cite{Renes2004}. Furthermore, the EJM is one of the few two-qubit measurements that can be efficiently localized, i.e., implemented via local operations at low entanglement cost~\cite{Pauwels2025}.

This symmetry and geometric structure give rise to several notable applications. In particular, it generates the so-called Elegant Distribution in the triangle network~\cite{Gisin2019}, which exhibits genuine network nonlocality—a form of nonclassical correlations irreducible to Bell nonlocality~\cite{Pozas2022,Gitton2024,VictorPirsa,Tavakoli2022review}. The EJM has also been used in bilocality tests~\cite{Tavakoli2021,Baeumer2021,Huang2022}, quantum teleportation~\cite{Ding2024}, and the study of measurement incompatibility~\cite{Patra2025}. These properties have established the EJM as a benchmark example of an entangled measurement. 

This motivates generalization—either to higher-dimensional systems or multipartite settings. Several attempts in this direction have been made and progress has been achieved in higher dimensions~\cite{Czartowski2021}. However, these efforts have faced persistent challenges: parametric extensions are very challenging, and some features do not generalize straightforwardly due to the complexity of multiparty entanglement~\cite{Ding2025}. Despite the interest it has attracted, a clean and operationally meaningful extension of the EJM to more than two subsystems has remained elusive.

Here, we take a different approach. We begin by abstracting the EJM as a special member of a family of locally encodable \cite{Tanaka007,Pimpel2023} bases characterized by tetrahedral symmetry, which we define for an arbitrary number of qubits. A detailed exploration of this family (including a higher-dimensional generalization) is presented in \cite{ltz4-vx18}. Here, we show how the EJM and its generalizations to more qubits emerge as efficiently localizable tetrahedral basis with regular tetrahedral structure.

To make this construction concrete, it is useful to give a schematic form for the multiqubit case. Each $n$-qubit EJM basis vector can be written as
\begin{equation} \label{eq:EJM-n}
\ket{\psi}
\;=\;
\sum_{\boldsymbol s\in\{\pm\}^{n}}
c_{\boldsymbol s}\,
\bigl|\,
s_1\,\BlochTet
\!\otimes\!
s_2\,\BlochTet
\!\otimes\!
\cdots
\!\otimes\!
s_n\,\BlochTet
\bigr\rangle,
\end{equation}
where each $s_l\in\{+,-\}$ selects the sign of the tetrahedral direction of qubit $l$, and ${\BlochTet}$ denotes one of the four tetrahedral directions $\vec{m}_l$. Crucially, while the tetrahedral directions vary across the basis, the coefficients $c_{\boldsymbol s}$ are fixed. Indeed, while a single state can always be written in this form, enforcing it simultaneously for an entire orthonormal basis is highly non-trivial.

\section{EJM as a group-orbit basis}

We begin by highlighting two key structural properties of the EJM.

First, the EJM is iso-entangled: all basis states have the same degree of entanglement. By the Schmidt decomposition, this implies that the basis is locally encodable~\cite{Tanaka007,Pimpel2023}; that is, each basis state can be obtained from a fixed fiducial state, which we may take to be $\ket{\psi_{\mathrm{EJM}}} \equiv \ket{\psi_{\mathrm{EJM},1}}$ in \eqref{eq:EJM}, via local unitaries. Explicitly, there exist qubit unitaries $U^{(1)}_j$ and $U^{(2)}_j$ such that $\ket{\psi_j} = U^{(1)}_j \otimes U^{(2)}_j \ket{\psi_{\mathrm{EJM}}}$.

Second, the EJM exhibits specific local symmetry: the reduced Bloch vectors of each qubit align with the vertices of a regular tetrahedron. This symmetry is generated by single-qubit Pauli operators $\langle X, Z \rangle$ acting on each qubit\footnote{%
Here, $\langle \cdot \rangle$ denotes the group generated under composition.%
}.

Together, these properties imply that the EJM forms a group-covariant orthonormal basis: the orbit of a single fiducial state under a finite symmetry group. Specifically, the abelian order-four group
\begin{equation}
G_{\mathrm{tetra}}^{(2)} \equiv \langle X \otimes X,\; Z \otimes Z \rangle\,,
\end{equation} generates the full basis from $\ket{\psi_{\mathrm{EJM}}}$.

Importantly, while the group $G_{\mathrm{tetra}}^{(2)}$ enforces local tetrahedral symmetry it does not specify its specific geometry: depending on the fiducial state, the local Bloch vectors may form any disphenoid (a specific type of tetrahedron with pairs of opposite edges of equal length) \cite{coxeter1973}, e.g. a regular tetrahedron, a distorted one, or a planar rectangle. In other words, while the group always enforces tetrahedral symmetry, it also stabilizes a variety of other local configurations and the actual geometry is co-determined by the fiducial state.

In what follows, we abstract away from the EJM itself and consider a family of orthonormal bases generated as group orbits of $G_{\mathrm{tetra}}^{(2)}$, to which we refer as tetrahedral bases. Our goal is to characterize all such bases and identify how the EJM arises as a distinguished member within this family i.e. as a regular tetrahedral basis.

\section{Qubit Tetrahedral Bases}

To define a natural \( n \)-qubit generalization of tetrahedral bases, we impose two structural requirements on its symmetry group: (i) local encodability via tetrahedral single-qubit orbits, implemented by the Pauli group \( \langle X, Z \rangle \), and (ii) a global abelian group structure acting on the full basis.

These constraints specify (up to local Clifford equivalence\footnote{That is, any other group satisfying these properties is related by Clifford conjugation.}) the symmetry group
\begin{equation}
G^{(n)}_{\mathrm{tetra}} \equiv \left\langle Z^{(i)} Z^{(i+1)},\, X^{\otimes n} \right\rangle,
\end{equation}
where $Z^{(i)}$ acts on qubit $i$, and $i$ runs from $1$ to $n-1$. The group is abelian of order $2^n$, and every Pauli operator appears locally.

We define a qubit tetrahedral basis as any orthonormal basis obtained from the orbit of a fiducial state $\ket{\psi}$ under the group $G^{(n)}_{\mathrm{tetra}}$:
\begin{equation}
\mathcal{B} = \bigl\{\, U_g \ket{\psi} \;\big|\; U_g \in G^{(n)}_{\mathrm{tetra}} \,\bigr\},
\end{equation}
with the additional requirement that $\langle \psi | U_g^\dagger U_{g'} | \psi \rangle = \delta_{g,g'}$.
Since $G^{(n)}_{\mathrm{tetra}}$ decomposes into $2^n$ one-dimensional irreducible representations, orthonormality holds whenever $\ket{\psi}$ is an equal-weight superposition over their eigenspaces. This follows directly from Schur's lemma \cite{ltz4-vx18}.

To construct these bases explicitly, we use that \( G^{(n)}_{\mathrm{tetra}} \) is isomorphic to \( \mathbb{Z}_2^n \) and can be generated by the commuting Pauli's
\begin{equation}
\langle Z^{(1)}, \dots, Z^{(n-1)}, X^{(n)} \rangle \;\cong\; G^{(n)}_{\mathrm{tetra}}.
\end{equation}
Its simultaneous eigenbasis is the product basis \( \ket{z_1,\dots,z_{n-1},x} \equiv \ket{\vec{z},x} \), with \( z_i,x \in \{0,1\} \).

To relate this representation to the representation \( G^{(n)}_{\mathrm{tetra}} \) defining the tetrahedral basis, we introduce the unitary
\begin{equation} \label{eq:cnots}
S(n) = \mathrm{CNOT}_{(2 \to 1)} \cdots \mathrm{CNOT}_{(n \to n-1)},
\end{equation}
where \( \mathrm{CNOT}_{(c \to t)} \) denotes a CNOT gate with control qubit \( c \) and target qubit \( t \).  
This circuit maps the product eigenbasis to one that diagonalizes \( G^{(n)}_{\mathrm{tetra}} \).

It follows that, for any choice of phases $\alpha_{\vec{z},x} \in \mathbb{R}$, the orbit of the state
\begin{equation} \label{eq:tetrabasis}
\ket{\psi} = \frac{1}{2^{n/2}} \sum_{\vec{z} \in \mathbb{Z}_2^{n-1},\, x \in \mathbb{Z}_2} e^{i \alpha_{\vec{z},x}}\, S(n) \ket{\vec{z},x}
\end{equation}
defines an orthonormal basis that is covariant under $G^{(n)}_{\mathrm{tetra}}$ and inherits local tetrahedral symmetry through the structure of the group action.

This construction yields a large family of symmetric multiqubit bases, parametrized by phase choices $\alpha_{\vec{z},x}$~\cite{ltz4-vx18}. The EJM corresponds to a specific element of this family characterized by reduced states whose Bloch vectors form a regular tetrahedron. Imposing this directly on the marginals leads to nonlinear equations which are unsolvable in general. Instead, we leverage a second defining feature of the EJM, its efficient localizability, as a more tractable and operationally meaningful criterion.

\section{EJM from Localizability}

The EJM is locally encodable: each basis state $\ket{\psi_i}$ can be generated from a single fiducial state via local unitaries. Localizability asks the complementary question: can the basis states be reliably distinguished using only local operations and shared entanglement \footnote{We note that for projective measurements, the ability to distinguish the basis states determinstically is equivalent to being able to implement the measurement itself. This follows from the linearity of the Born rule.}?

While this is always possible in principle~\cite{Vaidman2003,Groisman2003}, the amount of entanglement this requires may be very large or even infinite. The entanglement cost thus provides an operational measure of the nonlocal complexity of a measurement.

This was formalized in Ref.~\cite{Pauwels2025}, where it was shown that the entanglement cost of implementing a measurement basis via local operations depends on the membership of the associated measurement unitary
\begin{equation}
M = \sum_i \ket{\psi_i}\!\bra{i}\,,
\end{equation}
where $\{\ket{i}\}$ denotes the computational basis and $\{\ket{\psi_i}\}$ the measurement basis,
in a hierarchy of sets \(\mathcal{V}_1 \subsetneq \dots \subsetneq \mathcal{V}_{k-1} \subsetneq \mathcal{V}_k \subsetneq\dots\). The integer \(k \geq 1\) labels a level in this hierarchy and relates to the amount of entanglement required to localize the measurement. Specifically, the minimal entanglement cost is determined by the smallest \(k\) such that \(M \in \mathcal{V}_k\). The structure of $\mathcal{V}_k$ is poorly understood, and an explicit characterization is generally intractable. However, the EJM is known to be in the lowest level where partially entangled bases appear\footnote{We label levels starting from \(k = 1\) to align with the standard convention for the Clifford hierarchy.}: level $k=3$.

At first, the difficulty of characterizing \(\mathcal{V}_k\) explicitly seems to hinder the generalization to more subsystems. However, for tetrahedral bases, the additional symmetry imposes a strong structure that allows significant simplification~\cite{ltz4-vx18}.

A key insight is that the $k$th level $\mathcal{C}_k$ of the Clifford hierarchy~\cite{Gottesman1999}, a complexity-theoretic classification of unitaries, forms a strict subset of $\mathcal{V}_k$. While some localizable bases lie in $\mathcal{V}_k$ but not in $\mathcal{C}_k$, the EJM itself is known to belong to $\mathcal{C}_3\subsetneq \mathcal{V}_3$~\cite{Pauwels2025}.

First, we make the following observation on the structure of tetrahedral fiducials:
\begin{observation}[Fiducial normal form]
Every tetrahedral fiducial state can be prepared  as $\ket{\psi} = V\ket{0}^{\otimes n}$ using a unitary of the form
\begin{equation} \label{eq:GenFidPhase}
V = S(n)\, H_n\, D_{\vec \alpha}\, H^{\otimes n}\,,
\end{equation}
where $D_{\vec \alpha}$ is a diagonal phase gate (which contains the phases $e^{i \alpha_{\vec{z},x}}$ from Eq.~\eqref{eq:tetrabasis}), $H_n$ denotes the single-qubit Hadamard acting only on qubit $n$, whereas $H^{\otimes n}$ is the $n$-fold tensor product of Hadamards acting on all qubits, and $S(n)$ is the operator defined in Eq.~\eqref{eq:cnots}. 
\end{observation}

We now state the key theorem, proven in the companion work~\cite{ltz4-vx18}, that characterizes the localizability of tetrahedral bases.
\begin{theorem}[Clifford level of a tetrahedral measurement unitary]\label{lem:localizability}
Fix a fiducial state $\ket{\psi}$ and define the basis-change unitary
\begin{equation}
M_\psi \;:=\; \sum_{g:U_g\in G^{(n)}_{\mathrm{tetra}}} U_g\,\ket{\psi}\bra{g},
\end{equation}
with $\{\ket{g}\}$ denoting the computational basis, so that $M_\psi\ket g=U_g\ket{\psi}=\ket{\psi_g}$.
Considering the diagonal phase gate $D_{\vec \alpha}$ that appears in the normal form of Eq.~\eqref{eq:GenFidPhase}, one has, for $k\ge2$, that
\begin{equation}\label{eq:fiducial_determines_Clifford}
D_{\vec \alpha}\in\mathcal{C}_k \quad\Longrightarrow\  M_\psi \in \mathcal{C}_k.
\end{equation}
\end{theorem}

Remembering that $\mathcal{C}_k \subsetneq \mathcal{V}_k$, the theorem above implies that the localizability of a tetrahedral basis can be analyzed by examining the diagonal phase gate used to construct the fiducial.
Diagonal unitaries in the Clifford hierarchy were explicitly characterized by Ref.~\cite{Gottesman2017}. Any diagonal gate $D_f$ acting on $n$ qubits can be approximated as
\begin{equation} \label{eq:fDiag}
D_f \ket{\vec{z}} = \exp\left(2\pi i\, \frac{f(\vec{z})}{2^m}\right) \ket{\vec{z}} \,,
\end{equation}
where \[f: \mathbb{Z}_2^n \rightarrow \mathbb{Z}_{2^m}\] is an integer-valued polynomial and $m \in \mathbb{N}$ determines the phase precision.

The Clifford level of $D_f$ is jointly determined by the weighted degree and precision $m$ of the phase polynomial $f$~\cite{Gottesman2017}. Since $f$ is defined over a finite field and has bounded weighted degree, only finitely many such polynomials exist at each level $k$. In~\cite{ltz4-vx18}, we enumerate all admissible $D_f$ for two qubits and classify the corresponding tetrahedral bases. Here, we focus on those whose local reductions form a regular tetrahedron—i.e., multiqubit generalizations of the EJM.

For two qubits, our criteria of local regular tetrahedral symmetry combined with efficient localizability uniquely identify a single solution at Clifford level $k = 3$ \cite{Pauwels2025}. This corresponds to the phase polynomial
\[
f(z_1, z_2) = z_1 z_2 
\]
and $m=2$. Building the fiducial state through the circuit \eqref{eq:GenFidPhase}, one finds that the orbit of $G_\mathrm{tetra}^{(2)}$
corresponds to the original EJM defined in Eq.~\eqref{eq:EJM}, see Appendix~\ref{app:2qubit}. Although other configurations—such as rectangles or collinear arrangements—also arise at this level, the requirement that all Bloch components be non-zero uniquely selects this canonical structure.

\section{Three qubit EJM}
The situation becomes richer for three qubits. At Clifford level $k=3$, tetrahedral bases only extend trivial two-qubit geometries (e.g., a tetrahedron combined with a planar structure) \cite{ltz4-vx18}. However, at level $k=4$, fully regular tetrahedral configurations emerge, all with local Bloch vectors exactly half the norm of the two-qubit EJM, $|\vec{m}|=\sqrt{3}/4$. All configurations have the same geometric chirality: the local Bloch vectors form identically oriented tetrahedra on each qubit. This consistent handedness distinguishes them from the two-qubit EJM. Remarkably, despite identical local geometry, these give rise to multiple locally inequivalent bases. We identify four different classes based on the three-tangle of the fiducial state (quantifying genuine tripartite entanglement~\cite{Coffman2000}). For each tangle, we have two classes of bases which differ by complex conjugation (which cannot be implemented with local unitaries). All other bases are equivalent up to local Clifford unitaries.

All fiducials exhibit identical pairwise concurrence across all qubit pairs, underscoring their high symmetry and the genuinely tripartite nature of the entanglement, equally distributed among parties—unlike the parametric construction of~\cite{Ding2025}, which lacks genuine three-party entanglement. Full details and explicit representatives are provided in End Matter.

Explicitly, a representative of the least entangled class (with full party-permutation symmetry) arises from the phase polynomial  
\[
f(z_1,z_2,z_3) = 3 z_1 z_3 + z_2 z_3 + 3 z_1 z_2 z_3 \,,
\]
and precision $m=2$.
The corresponding fiducial state is then constructed via the circuit  Eq.~\eqref{eq:GenFidPhase}. 
This state has all reduced Bloch vectors point along $\vec{m}_1$, which allows us to express it as
\begin{equation}
\begin{aligned}
\ket{\psi} =\;& 
c_0\, \ket{-\vec{m}_1,-\vec{m}_1,-\vec{m}_1} \\
&+ c_1\, \big( \ket{\vec{m}_1,-\vec{m}_1,-\vec{m}_1} + \text{perms.} \big) \\
&+ c_2\, \big( \ket{\vec{m}_1,\vec{m}_1,-\vec{m}_1} + \text{perms.} \big) \\
&+ c_3\, \ket{\vec{m}_1,\vec{m}_1,\vec{m}_1} \,.
\end{aligned}
\end{equation}
This makes explicit how the action of $G^{(3)}_{\mathrm{tetra}}$ generates a full orthonormal basis with regular tetrahedral structure, as in Eq.~\eqref{eq:EJM-n}. The full derivation and analytic expressions for the coefficients $c_i$ are provided in Appendix~\ref{app:tetrabasis}.

\section{Four qubits and beyond}

For four qubits, we find many regular tetrahedral basis configurations, which first appear at level~$k=5$ of the Clifford hierarchy. Compared to the two- and three-qubit cases, two new features emerge. First, we observe the existence of multiple geometric chiralities. Second, we identify two different tetrahedron sizes: one with $\lvert\vec{m}\rvert = \sqrt{3}/8$, which is exactly half the size of the two-qubit case, and another with $\lvert\vec{m}\rvert = 3\sqrt{3}/8$, which lies between the sizes observed in the two- and three-qubit cases. We provide two explicit examples in Appendix~\ref{app:4qubitexamples}. 

These bases realize genuine multiqubit extensions of the EJM, demonstrating that such structures persist in larger systems. A full classification of four-qubit (and higher) tetrahedral bases is left for future work.

These observations motivate the following conjecture.

\begin{conjecture}
Regular tetrahedral measurement bases, such as the Elegant Joint Measurement (EJM), exist for arbitrary \( n \) and always lie in level \( \mathcal{C}_{n+1} \) of the Clifford hierarchy with precision $m=2$. While multiple inequivalent EJM classes emerge as \( n \) grows, we conjecture that at least one such class gives rise to a family in which the reduced Bloch vectors exhibit predicable length scaling, namely \( \| \vec{m} \| = \sqrt{3}/2^{n-1} \).
\end{conjecture}

\section{Discussion}

This work reveals a deep structure of some $n$-qubit joint measurements. All eigenstates share the same degree of partial entanglement and all local Bloch vectors point toward the vertices of the regular tetrahedron. Consequently, these  measurements demonstrate a combination of local and non-local structure. They build on a striking interplay between symmetry, group structure, local encodability, and localizability. By examining localizability within a broad class of group-covariant bases with tetrahedral symmetry, we identify the EJM as a distinguished, highly symmetric solution in this landscape. While the local geometry is uniquely fixed—each qubit exhibiting the same tetrahedral structure—the three-qubit case admits multiple inequivalent classes of globally entangled bases. All representatives display genuine tripartite entanglement and identical pairwise concurrences, highlighting the richer structure of multipartite entanglement.

These symmetric joint measurements are particularly well suited for quantum-network applications. In networks, joint measurements play the role of effective ``links'' between independently prepared sources, and imposing symmetry at the measurement level often induces highly constrained, analyzable correlation structures~\cite{Gisin2019,Gitton2024}. The EJM itself occupies a useful ``sweet spot'': it is just beyond Clifford (hence capable of enabling genuinely nonclassical behavior in settings where purely Clifford components cannot), yet retains a low-complexity, highly symmetric description. This makes the multipartite generalizations identified here natural candidates to explore extensions of bilocal scenarios (where an EJM-like measurement sits at a central station~\cite{Tavakoli2021}) to star networks with three or more branches, as well as fully connected ``elegant'' network constructions. In Appendix~\ref{app:network-correlations} we illustrate this by computing the distribution generated by a tetrahedral network in which each node performs a three-party EJM-type measurement; the resulting correlations inherit strong symmetry constraints, providing a concrete starting point for questions about genuine and full network nonlocality beyond bipartite links.

Relatedly, EJM-type joint measurements have been proposed as resources in modified teleportation protocols~\cite{Ding2024}; our multipartite EJM-type constructions naturally raise the question whether analogous \emph{multipartite} teleportation/broadcasting primitives exist, and what symmetry/robustness advantages they may offer. Finally, the two-qubit EJM was essentially present already in the context of optimally extracting a spatial direction encoded in entangled spin systems, where measurements tailored to antiparallel (rather than parallel) spins improve direction estimation~\cite{Massar1995,Gisin1999}; it would be interesting to investigate whether the multiqubit EJM-type measurements identified here offer advantages for decoding spatial-direction encodings from larger entangled spin ensembles, complementing earlier approaches based on covariant measurements and collective-spin methods~\cite{Bagan2000,Bagan2001}.

A central open question is whether there exists a closed-form expression for the phase polynomial defining the $n$-qubit EJM for arbitrary $n$. Such a result would complete the geometric and operational characterization initiated here. Going further, it would be interesting to find a general parametric family of EJM bases that interpolate between different entanglement classes, similar to the results of \cite{Tavakoli2021} in the two-qubit case.

\onecolumngrid

\section*{End matter}

\subsection*{Full Classification of Three-Qubit EJMs} \label{app:classification}

Every three-qubit Elegant Joint Measurement (EJM) basis is generated by a fiducial vector $\ket{\psi}$ such that all single-qubit reduced density matrices have Bloch vectors of identical length $|\vec{m}| = \sqrt{3}/4$ and directions pointing to the vertices of a regular tetrahedron. Despite this isotropy, the family of admissible fiducials is rich, and we classify all distinct EJM bases up to local unitaries (LU).

For each fiducial state, we compute two LU invariants:
\begin{itemize}
\item \textbf{Three-tangle}
    \[
    \tau(\psi) = 4|d_1 - 2d_2 + 4d_3|
    \]
    where
    \begin{align*}
    d_1 &= a_{000}^2 a_{111}^2 + a_{001}^2 a_{110}^2 + a_{010}^2 a_{101}^2 + a_{100}^2 a_{011}^2,\\
    d_2 &= a_{000} a_{111}(a_{011} a_{100} + a_{101} a_{010} + a_{110} a_{001})\\
         &\quad + a_{011} a_{100} a_{101} a_{010} + a_{011} a_{100} a_{110} a_{001} + a_{101} a_{010} a_{110} a_{001},\\
    d_3 &= a_{000} a_{110} a_{101} a_{011} + a_{111} a_{001} a_{010} a_{100},
    \end{align*}
    for $\ket{\psi} = \sum_{ijk} a_{ijk} \ket{ijk}$.

The three-tangle quantifies the genuine multiparty entanglement content of the state \cite{Coffman2000}.

\item \textbf{Pairwise concurrence squared} $(C^2_{AB}, C^2_{AC}, C^2_{BC})$: For a pair of qubits, concurrence for pure three-qubit state $\ket{\psi}$ is given by
    \[
    C_{AB} = \max\left(0, \sqrt{\lambda_1} - \sqrt{\lambda_2} - \sqrt{\lambda_3} - \sqrt{\lambda_4} \right),
    \]
    where $\lambda_i$ are the (sorted) eigenvalues of $\rho_{AB} \tilde{\rho}_{AB}$, with $\rho_{AB} = \mathrm{Tr}_C \ket{\psi}\bra{\psi}$ and $\tilde{\rho}_{AB}$ is the spin-flipped matrix. The triple $(C^2_{AB}, C^2_{AC}, C^2_{BC})$ uniquely labels the pairwise entanglement content.
\end{itemize}

Remarkably, we observe that within each three-tangle class, all pairwise squared concurrences—$C_{AB}^2$, $C_{AC}^2$, and $C_{BC}^2$—are identical. This uniformity underscores the highly symmetric nature of these EJM fiducials: not only are the single-qubit states isotropic, but the bipartite entanglement is also uniformly distributed across all qubit pairs. This also implies we only have one genuine LU invariant (the three-tangle). 

Within each class of fiducial states sharing the same three-tangle, we find that all representatives are equivalent up to local Clifford operations and complex conjugation in the computational basis. The latter operation  cannot be implemented by local unitaries alone and can be interpreted as inducing a distinct chirality of the entanglement.

Since $G^{(3)}_\mathrm{tetra}$ is an Abelian subgroup of the Pauli group, the tetrahedral basis it generates is preserved under all local Clifford operations. Consequently, local Clifford transformations map the EJM basis to itself (up to column permutations and overall phases), ensuring equivalence at the level of the bases.

To further characterize the structure of each class of tetrahedral bases, we analyze the symmetry of the fiducial states under qubit permutations. For each representative state, we compute the order of its stabilizer subgroup within the symmetric group \( S_3 \), defined as the set of qubit permutations that leave the state invariant (up to global phase). This provides a measure of the inherent permutation symmetry of the state.

We find that only the fiducials in the lowest-tangle class (with \( \tau = \sqrt{65}/16 \)) exhibit full permutation symmetry, i.e., they are invariant under all six elements of \( S_3 \). In contrast, the fiducials in the remaining classes are invariant under at most a single transposition, corresponding to a stabilizer subgroup of order 2. 

Finally, we note that all fiducials generate tetrahedra with the same geometric chirality: the local tetrahedra on each qubit have the same handed orientation and are not mirror images of each other. This contrasts with the two-qubit EJM, where the local tetrahedra have opposite chirality.

This geometric chirality is unrelated to entanglement chirality, which arises from complex conjugation. Conjugation flips the sign of the Y component for all qubits and cannot be implemented by local unitaries. It leads to distinct, locally inequivalent fiducials with identical entanglement measures. The geometric chirality of the basis is invariant under local unitaries, including local Cliffords. 

Table~\ref{table:classes} summarizes these results. The code used for this analysis can be found in \cite{compapp}.

\begin{table}[h!]
\centering
\renewcommand{\arraystretch}{1.3}
\begin{tabular}{|c|c|c|c|c|}
\hline
\textbf{$\tau$} & \textbf{ \( C^2\)} & \textbf{Fiducial} & \textbf{Phase Polynomial} & $|S_3|$ \\
\hline
\( \frac{\sqrt{65}}{16} \) & 
\(\frac{1}{32} \left(13-\sqrt{65}\right) \) &
\(\frac{1}{2}\,[\,1,\; \tfrac{1}{2}(1+i),\; \tfrac{1}{2}(1+i),\; \tfrac{1}{2}(1{-}i),\; \tfrac{1}{2}(1+i),\; \tfrac{1}{2}(1{-}i),\; \tfrac{1}{2}(1{-}i),\; 0\,]\) & \( z_1 z_3 + 3 z_2 z_3 + z_1 z_2 z_3 \) & 1,2,6 \\
& & 
\(\frac{1}{2}\,[\,1,\; \tfrac{1}{2}(1{-}i),\; \tfrac{1}{2}(1{-}i),\; \tfrac{1}{2}(1{+}i),\; \tfrac{1}{2}(1{-}i),\; \tfrac{1}{2}(1{+}i),\; \tfrac{1}{2}(1{+}i),\; 0\,]\) & \( 3 z_1 z_3 + z_2 z_3 + 3 z_1 z_2 z_3 \) & \\
\hline
\( \frac{\sqrt{97}}{16} \) & 
$\frac{1}{32} \left(13-\sqrt{97}\right)$ &
\(\frac{1}{2}\,[\,1,\; \tfrac{1}{2}(1{-}i),\; -\tfrac{1}{2}(1{-}i),\; \tfrac{1}{2}(1{-}i),\; \tfrac{1}{2}(1{+}i),\; \tfrac{1}{2}(1{+}i),\; \tfrac{1}{2}(1{+}i),\; 0\,]\) & \( z_1 z_2 + z_1 z_3 + z_2 z_3 + 3 z_1 z_2 z_3 \) & 1,2\\
& & 
\(\frac{1}{2}\,[\,1,\; \tfrac{1}{2}(1{+}i),\; -\tfrac{1}{2}(1{-}i),\; \tfrac{1}{2}(1{-}i),\; \tfrac{1}{2}(1{+}i),\; \tfrac{1}{2}(1{+}i),\; \tfrac{1}{2}(1{-}i),\; 0\,]\) & \( z_1 z_2 + z_1 z_3 + 3 z_2 z_3 + z_1 z_2 z_3 \) &  \\
\hline
\( \frac{\sqrt{113}}{16} \) & 
$\frac{1}{32} \left(13-\sqrt{113}\right)$ &
\(\frac{1}{2}\,[\,1,\; 0,\; \tfrac{1}{2}(1{+}i),\; 1,\; 0,\; -\tfrac{1}{2}(1{-}i),\; 1,\; 0\,]\) & \( z_1 z_2 + 2 z_1 z_3 + z_1 z_2 z_3 \)& 1,2 \\
& & 
\(\frac{1}{2}\,[\,1,\; 0,\; -\tfrac{1}{2}(1{-}i),\; 1,\; 0,\; -\tfrac{1}{2}(1{+}i),\; 1,\; 0\,]\) & \( 2 z_1 z_2 + 2 z_1 z_3 + z_1 z_2 z_3 \) & \\
\hline
\( \frac{\sqrt{145}}{16} \) & 
$\frac{1}{32} \left(13-\sqrt{145}\right)$ &
\(\frac{1}{2}\,[\,1,\; 0,\; 0,\; \tfrac{1}{2}(1{-}i),\; \tfrac{1}{2}(1{+}i),\; 1,\; 1,\; 0\,]\) & \( z_1 z_3 + z_1 z_2 z_3 \) & 1,2 \\
& & 
\(\frac{1}{2}\,[\,1,\; 0,\; \tfrac{1}{2}(1{-}i),\; 1,\; 0,\; \tfrac{1}{2}(1{+}i),\; 1,\; 0\,]\) & \( 2 z_1 z_3 + z_1 z_2 z_3 \) & \\
\hline
\end{tabular}
\caption{One representative fiducial for each inequivalent EJM class, grouped by 3-tangle.
Within each tangle class, all pairwise concurrences are equal: $C^2 = C_{AB}^2 = C_{BC}^2 = C_{AC}^2$.
Each class contains two locally inequivalent chiralities related by complex conjugation (followed by local Clifford unitaries).
Fiducials are expressed in the computational basis, and all phase polynomials are defined with precision $m = 2$ (all bases that result from phase polynomials with higher precision are equivalent to a basis with $m=2$).
The last column indicates the possible sizes of the stabilizer subgroup of $S_3$ within each class.
\label{table:classes} }
\end{table}

\paragraph*{Bloch vector alignment via local unitaries.} 
The Bloch vectors of the single-qubit reduced states can be rotated via local unitaries without changing the entangled structure of the three-qubit fiducial. For example, for the first fiducial state listed in Table~\ref{table:classes}, one can align all three Bloch vectors to point in the direction of \( \vec{m}_1 \) by applying the local unitary:

\[
U = \frac{-i}{\sqrt{2}}\, (Z_1 - X_1)
\cdot \frac{1}{\sqrt{2}}\, (Z_2 - X_2)
\cdot \frac{1}{\sqrt{2}}\, (Z_3 - X_3)
\cdot \bigotimes_{j=1}^3 \exp\left( \frac{i \pi}{3} \cdot \frac{X_j - Y_j - Z_j}{\sqrt{3}} \right).
\]
Applying this unitary to the fiducial state yields the normalized vector
\[
\ket{\psi} = \frac{1}{\sqrt{8}} \left( -1 - i,\; 1,\; 1,\; 1,\; 1,\; 1,\; 1,\; 0 \right),
\]
in the computational basis.

Note that the single-qubit unitary
$
\frac{Z - X}{\sqrt{2}}
$
maps each Bloch vector direction \( \vec{m}_j \) to \( -\vec{m}_{\pi(j)} \), where the permutation \( \pi \) acts as \( \pi = (1\;2\;4\;3) \).

\begin{acknowledgments}
We thank Cyril Branciard, Alejandro Pozas-Kerstjens, Matthias Christandl, Teiko Heinosaari, Pauli Jokinen and Pavel Sekatski for discussions. 
We acknowledge financial support from Swiss National Science Foundations (NCCR-SwissMAP).
\end{acknowledgments}

\bibliography{refs}

\appendix
\input{appendices}

\end{document}

%% file: appendices.tex
\section{Two-qubit EJM} \label{app:2qubit}

 The two-qubit EJM is defined as the orthonormal basis
\begin{equation}
\ket{\psi_i}= \frac{\sqrt{3}+1}{2\sqrt{2}}\ket{\vec{m}_i,-\vec{m}_i}+\frac{\sqrt{3}-1}{2\sqrt{2}}\ket{-\vec{m}_i,\vec{m}_i}\,,
\end{equation}
where \( \ket{\vec{m}} = \cos\left(\tfrac{\theta}{2}\right)\ket{0} + \sin\left(\tfrac{\theta}{2}\right) e^{\ii \phi} \ket{1} \) and \( \ket{-\vec{m}} = \sin\left(\tfrac{\theta}{2}\right)\ket{0} - \cos\left(\tfrac{\theta}{2}\right) e^{\ii \phi} \ket{1} \), with \( \theta = \arccos(m_z) \), \( \phi = \arg(m_x + \ii m_y) \), and \( \vec{m} = (m_x, m_y, m_z) \in \mathbb{R}^3 \) along Bloch vector \( \vec{m}_i \). 

In the computational basis, we can write the basis as
\begin{equation} \label{eq:EJMmatrix}
  M = \left(
\begin{array}{cccc}
 \frac{1}{2}+\frac{i}{2} & -\frac{1}{2}+\frac{i}{2} & \frac{1}{2}-\frac{i}{2} &
   -\frac{1}{2}-\frac{i}{2} \\
 -i & 0 & 0 & -i \\
 0 & i & i & 0 \\
 \frac{1}{2}-\frac{i}{2} & -\frac{1}{2}-\frac{i}{2} & \frac{1}{2}+\frac{i}{2} &
   -\frac{1}{2}+\frac{i}{2} \\
\end{array}
\right)\,
\end{equation}
where each column corresponds to a basis vector.

From the phase polynomial $f(z_1,z_2) = z_1z_2$, we build the fiducial state via the circuit of Eq.~\eqref{eq:GenFidPhase}. Explicitly, 

\begin{align}
 \ket{\psi} &= \mathrm{CNOT}_{(2 \to 1)} H_2 \mathrm{diag}(1,1,1,e^{\ii 2 \pi/4})\ket{+,+} \\ &=  [1,
 \frac{1}{2}-\frac{i}{2}, 
 \frac{1}{2}+\frac{i}{2},
 0 ]^T/\sqrt{2} \,,
\end{align}
which corresponds to the first basis state of \eqref{eq:EJMmatrix} up to the local unitary $-Y_2\ket{\psi}$.

\section{Analytic Decomposition in the Tetrahedral Product Basis} \label{app:tetrabasis}

Here, we derive the analytic form of the most symmetric three-qubit tetrahedral EJM fiducial state. This state arises from the phase polynomial
\[
f(z) = 3z_1 z_3 + z_2 z_3 + 3 z_1 z_2 z_3 \mod 4\,,
\]
via the circuit construction of Eq.~\eqref{eq:GenFidPhase}.

Explicitly, we find
\begin{align}
\ket{\psi} &= V\ket{0}^{\otimes 3}\\  &= CNOT_{2\rightarrow 1} CNOT_{3\rightarrow 2} H_3 \sum_{\vec{z}} \exp\left(2\pi i\, \frac{f(\vec{z})}{2^m}\right)\ket{\vec{z}} \bra{\vec{z}}  H^{\otimes 3}  \ket{0}^{\otimes 3} \\ &= 
\frac{1}{2}\begin{bmatrix}
 1 & 
\frac{1}{2} - \frac{i}{2} & 
\frac{1}{2} - \frac{i}{2} & 
\frac{1}{2} + \frac{i}{2} & 
\frac{1}{2} - \frac{i}{2} & 
\frac{1}{2} + \frac{i}{2} & 
\frac{1}{2} + \frac{i}{2} & 
0
\end{bmatrix}^T.
\end{align}
It is instructive to express this in the product basis $\{\ket{\pm\vec{m}_1}\}^{\otimes 3}$, where $\vec{m}_1 = (1,1,1)/\sqrt{3}$. Performing the basis transformation, we find
\[
\begin{aligned}
\ket{\psi} = \frac{1}{12}\bigg[ \;
& \gamma_+ e^{i \theta_0} \,\ket{+\vec{m}_1,+\vec{m}_1,+\vec{m}_1} \\
& + \delta_+\, e^{i\theta_1} \left(
\ket{+\vec{m}_1,+\vec{m}_1,-\vec{m}_1}
+ \ket{+\vec{m}_1,-\vec{m}_1,+\vec{m}_1}
+ \ket{-\vec{m}_1,+\vec{m}_1,+\vec{m}_1}
\right) \\
& + \delta_-\, e^{i\theta_2} \left(
\ket{+\vec{m}_1,-\vec{m}_1,-\vec{m}_1}
+ \ket{-\vec{m}_1,+\vec{m}_1,-\vec{m}_1}
+ \ket{-\vec{m}_1,-\vec{m}_1,+\vec{m}_1}
\right) \\
& + \gamma_-\, e^{i\theta_3} \ket{-\vec{m}_1,-\vec{m}_1,-\vec{m}_1} \;
\bigg]
\end{aligned}
\]
where:
\begin{align}
    \gamma_\pm &= \sqrt{45 \pm 17\sqrt{3}}\,, \\
    \delta_\pm &= \sqrt{9 \pm \sqrt{3}}\,,
\end{align}

and the phases
\begin{align}
    \theta_0 &= -\arctan(A_+)\,, \\
    \theta_1 &= \arctan(B_+)\,, \\
    \theta_2 &= \arctan(B_-)\,, \\
    \theta_3 &= \arctan(A_-)\,,
\end{align}
where $A_\pm = \frac{3}{37}(\pm 8 + 3\sqrt{3})$ and $
    B_\pm = 3(2 \pm \sqrt{3})$.

This decomposition makes explicit that the local action of $G_\text{tetra}^{(3)}$ leads to a basis of the general form Eq.~\eqref{eq:EJM-n}.

\section{Four qubit EJM examples} \label{app:4qubitexamples}

For four qubits, we again find many fiducial states generating regular tetrahedral measurement configurations. A full classification is left for future work. Here, we give two explicit examples that illustrate features not present in the three-qubit case.

First, we find two distinct tetrahedron sizes, characterized by the magnitude of the single-qubit Bloch vectors. One class has $\lvert\vec{m}\rvert = \sqrt{3}/8$, precisely half the size of the three-qubit case (and one quarter of the two-qubit case). The second class has $\lvert\vec{m}\rvert = 3\sqrt{3}/8$, which lies in between the two- and three-qubit values.

Second, while all three-qubit solutions have the same geometric chirality, four-qubit configurations can exhibit different geometric chiralities. That is, the local Bloch tetrahedra on different qubits can be mirror images of each other.

\subsection*{Example 1: Fully PPI, isotropic with minimal Bloch vector length}

At precision $m = 2$, the following phase polynomial defines a fiducial state in the regular tetrahedral configuration:
\[
\begin{aligned}
f(z) =\, & z_1 z_3 + z_1 z_4 + z_2 z_3 + 3 z_3 z_4 \\
        & + z_1 z_2 z_4 + z_1 z_3 z_4 + z_2 z_3 z_4 + 3 z_1 z_2 z_3 z_4 \mod 4.
\end{aligned}
\]
This leads to the fiducial state
\begin{align*}
\lvert\psi\rangle = \frac{1}{2\sqrt{2}} \Bigl(
&\ket{0000}
+\tfrac{1+i}{2}\,\ket{0001}
+i\,\ket{0010}
-\ket{0101}
+\tfrac{-1+i}{2}\,\ket{0110}
+\tfrac{1-i}{2}\,\ket{0111} \\
&+\tfrac{1+i}{2}\,\ket{1000}
+\tfrac{1+i}{2}\,\ket{1001}
+\ket{1011}
+\ket{1100}
+\tfrac{1-i}{2}\,\ket{1110}
\Bigr).
\end{align*}
This state is fully party-permutation-invariant (PPI), with all Bloch vectors aligned along the $\vec{m}_1$ axis,
\[\bigl(\langle X_i\rangle,\langle Y_i\rangle,\langle Z_i\rangle\bigr) = \left(\tfrac{1}{8},\, \tfrac{1}{8},\, \tfrac{1}{8}\right)\]
and length $\lvert\vec{m}\rvert = \sqrt{3}/8$.

\subsection*{Example 2: Chirality and larger tetrahedron size}

At precision $m = 2$, the phase polynomial
\[
f(z) = z_2 z_3 + 3 z_3 z_4 + 2 z_1 z_2 z_3 + z_1 z_2 z_4 + 3 z_1 z_3 z_4 + z_1 z_2 z_3 z_4 \mod 4
\]
produces the fiducial state
\begin{align*}
\lvert\psi\rangle = \frac{1}{2\sqrt{2}} \Bigl(
&\ket{0000}
+\tfrac{1+i}{2}\,\ket{0001}
+\tfrac{1+i}{2}\,\ket{0010}
+\tfrac{1+i}{2}\,\ket{0100}
+\ket{1000}
+\ket{1001}
- i\,\ket{1010} \\
&+\tfrac{1 - i}{2}\,\ket{1011}
+\ket{1100}
+\tfrac{-1 + i}{2}\,\ket{1101}
+\tfrac{1 - i}{2}\,\ket{1110}
\Bigr).
\end{align*}

This state has local Bloch vectors:
\[
\begin{aligned}
\bigl(\langle X_1\rangle,\langle Y_1\rangle,\langle Z_1\rangle\bigr) &= \left(\tfrac{3}{8},\, -\tfrac{3}{8},\, -\tfrac{3}{8}\right),\\
\bigl(\langle X_2\rangle,\langle Y_2\rangle,\langle Z_2\rangle\bigr) &= \left(\tfrac{3}{8},\, \tfrac{3}{8},\, \tfrac{3}{8}\right),\\
\bigl(\langle X_3\rangle,\langle Y_3\rangle,\langle Z_3\rangle\bigr) &= \left(\tfrac{3}{8},\, -\tfrac{3}{8},\, \tfrac{3}{8}\right),\\
\bigl(\langle X_4\rangle,\langle Y_4\rangle,\langle Z_4\rangle\bigr) &= \left(\tfrac{3}{8},\, \tfrac{3}{8},\, \tfrac{3}{8}\right).
\end{aligned}
\]
From the sign pattern, we observe that under the action of the group \( G_{\text{tetra}}^{(4)} \), this fiducial generates a tetrahedron where qubits 1, 2, and 4 point along a common tetrahedron (aligned with the \( (1,1,1) \) direction), while qubit 3 has a mirrored configuration.

\section{Tetrahedral Network Correlations} \label{app:network-correlations}

\begin{figure}[h!]
  \centering
  \begin{tikzpicture}[scale=1.1, every node/.style={font=\small}]

    \node[draw, thick, rounded corners=2pt, minimum width=0.64cm, minimum height=0.48cm] (A) at (-2,1.25) {$A$};
    \node[draw, thick, rounded corners=2pt, minimum width=0.64cm, minimum height=0.48cm] (B) at ( 2,1.25) {$B$};
    \node[draw, thick, rounded corners=2pt, minimum width=0.64cm, minimum height=0.48cm] (C) at ( 0,-2.2) {$C$};
    \node[draw, thick, rounded corners=2pt, minimum width=0.64cm, minimum height=0.48cm] (D) at ( 0,0) {$D$};

    \foreach \p/\q in {A/B, B/C, C/A, A/D, B/D, C/D} {
      \draw[very thick] (\p) -- (\q);
      \node[fill=white, inner sep=1pt] at ($(\p)!0.5!(\q)$) {$\psi^-$};
    }

    \node[draw,circle,inner sep=1pt, font=\scriptsize, fill=white] at ($(A)!0.28!(B)$) {1};
    \node[draw,circle,inner sep=1pt, font=\scriptsize, fill=white] at ($(A)!0.32!(D)$) {2};
    \node[draw,circle,inner sep=1pt, font=\scriptsize, fill=white] at ($(A)!0.28!(C)$) {3};
    \node[draw,circle,inner sep=1pt, font=\scriptsize, fill=white] at ($(B)!0.28!(A)$) {1};
    \node[draw,circle,inner sep=1pt, font=\scriptsize, fill=white] at ($(B)!0.28!(C)$) {2};
    \node[draw,circle,inner sep=1pt, font=\scriptsize, fill=white] at ($(B)!0.32!(D)$) {3};
    \node[draw,circle,inner sep=1pt, font=\scriptsize, fill=white] at ($(C)!0.28!(A)$) {1};
    \node[draw,circle,inner sep=1pt, font=\scriptsize, fill=white] at ($(C)!0.32!(D)$) {2};
    \node[draw,circle,inner sep=1pt, font=\scriptsize, fill=white] at ($(C)!0.28!(B)$) {3};
    \node[draw,circle,inner sep=1pt, font=\scriptsize, fill=white] at ($(D)!0.24!(A)$) {1};
    \node[draw,circle,inner sep=1pt, font=\scriptsize, fill=white] at ($(D)!0.24!(B)$) {2};
    \node[draw,circle,inner sep=1pt, font=\scriptsize, fill=white] at ($(D)!0.24!(C)$) {3};

  \end{tikzpicture}
  \caption{Fully connected $K_4$ network drawn as a triangle $ABC$ with a central node $D$. Each edge carries a singlet $\ket{\psi^-}$. Numbers give the wire order for the three-qubit measurement $M$: $A=(AB,AD,AC)$, $B=(BA,BC,BD)$, $C=(CA,CD,CB)$, $D=(DA,DB,DC)$. All parties apply the same three-qubit EJM in that order.}
  \label{fig:tetra-network}
\end{figure}

We consider the fully connected $K_4$ network (Fig.~\ref{fig:tetra-network}): six singlet sources
$\ket{\psi^-}$ connect every pair of parties $(A,B,C,D)$, so each party receives three qubits
(labeled by the incident edges, e.g.\ $A$ receives $AB,AD,AC$). Each party performs the same
three-qubit EJM defined as the orbit of the fiducial $\tau_{65}$ (see Table~\ref{table:classes} in the End Matter) under the commuting subgroup
$\langle ZZI,\, IZZ,\, XXX\rangle$ with the fixed (right-handed) local wire order shown in Fig.~\ref{fig:tetra-network}, namely $A=(AB,AD,AC)$, $B=(BA,BC,BD)$, $C=(CA,CD,CB)$, and $D=(DA,DB,DC)$. This choice ensures that all parties implement the same (right-handed) three-qubit EJM.
(Reversing the local wire order gives a locally equivalent
implementation.)

We fix the outcome labeling by identifying each outcome $j\in\{1,\dots,8\}$ with the bit string
\[
1\!\leftrightarrow\!000,\;
2\!\leftrightarrow\!100,\;
3\!\leftrightarrow\!010,\;
4\!\leftrightarrow\!001,\;
5\!\leftrightarrow\!110,\;
6\!\leftrightarrow\!101,\;
7\!\leftrightarrow\!011,\;
8\!\leftrightarrow\!111,
\]
and setting $\ket{\psi_j}=(ZZI)^{b_1}(IZZ)^{b_2}(XXX)^{b_3}\ket{\tau_{65}}$ for the associated bits
$(b_1,b_2,b_3)$. The resulting probability distribution $p(a,b,c,d)$ ($a,b,c,d\in\{1,\dots,8\}$)
can be viewed as a multipartite analogue of the Elegant Distribution in the triangle~\cite{Gisin2019}.

A direct evaluation (see \texttt{k4\_dist.jl} in \cite{compapp}) shows that $p$ is remarkably sparse
and arithmetic: among the $4096$ entries, exactly $1568$ are zero, and the $2528$ nonzero entries
take only seven distinct dyadic values, 
\[
\Bigl\{\tfrac{1}{16384},\tfrac{1}{4096},\tfrac{5}{16384},\tfrac{1}{2048},\tfrac{17}{16384},\tfrac{5}{4096},\tfrac{1}{512}\Bigr\},
\]
with multiplicities $(576,288,1152,192,64,160,96)$, respectively. 

We also consider local outcome relabelings: a symmetry is a 4-tuple of permutations
$(m_A,m_B,m_C,m_D)$ of $\{1,\dots,8\}$ such that
\begin{equation}
p(a,b,c,d)=p\!\bigl(m_A(a),m_B(b),m_C(c),m_D(d)\bigr)\qquad\forall\,a,b,c,d.
\end{equation}
For the $K_4$ network with the $\tau_{65}$ three-qubit EJM at each node (and the fixed wiring of
Fig.~\ref{fig:tetra-network}), the symmetry group of such outcome relabelings has size $16$ and is
generated by the four commuting involutions in Table~\ref{tab:symmetries}. These symmetries are
inherited from the measurement construction: $ZZI$, $IZZ$, and $XXX$ permute the local EJM basis
(and hence relabel outcomes), and each singlet edge is unchanged under applying the same Pauli on
both of its qubits (up to a global phase). In contrast, no nontrivial permutation of the parties
leaves $p$ invariant for this fixed wiring convention.

\begin{table}[h!]
  \centering
  \renewcommand{\arraystretch}{1.15}
  \caption{Generators of the local outcome-relabeling symmetry group of $p(a,b,c,d)$ for $\tau_{65}$
  (fixed wiring of Fig.~\ref{fig:tetra-network}). Each row gives a relabeling
  $(m_A,m_B,m_C,m_D)$ such that $p(a,b,c,d)=p(m_A(a),m_B(b),m_C(c),m_D(d))$.}
  \label{tab:symmetries}
  {\begin{tabular}{c|cccc}
    \toprule
      & $A$ & $B$ & $C$ & $D$ \\
    \midrule
    $g_1$ & $M_{ZZI}$ & $M_{ZZI}$ & $M_{IZZ}$ & $M_{ZIZ}$ \\
    $g_2$ & $M_{IZZ}$ & $\mathrm{id}$ & $M_{ZZI}$ & $M_{ZIZ}$ \\
    $g_3$ & $M_{XXX}$ & $M_{XXX}$ & $M_{XXX}$ & $M_{XXX}$ \\
    $g_4$ & $\mathrm{id}$ & $M_{IZZ}$ & $M_{IZZ}$ & $M_{IZZ}$ \\
    \bottomrule
  \end{tabular}}

  \vspace{0.35em}
  \begin{minipage}{0.95\linewidth}\footnotesize
  In the outcome convention $j\leftrightarrow(b_1,b_2,b_3)$ above, the three basic relabelings are
  simply bit-flips: $M_{ZZI}$ flips $b_1$, $M_{IZZ}$ flips $b_2$, and $M_{XXX}$ flips $b_3$. We also use
  $M_{ZIZ}:=M_{ZZI}\circ M_{IZZ}$, which flips both $b_1$ and $b_2$.
  Equivalently, in cycle notation on $\{1,\dots,8\}$,
  \[
    M_{ZZI}=(1\,2)(3\,5)(4\,6)(7\,8),\quad
    M_{IZZ}=(1\,3)(2\,5)(4\,7)(6\,8),\quad
    M_{XXX}=(1\,4)(2\,6)(3\,7)(5\,8),\quad
    M_{ZIZ}=(1\,5)(2\,3)(4\,8)(6\,7).
  \]
  The generators $g_1,\dots,g_4$ commute and generate a $16$-element subgroup of local outcome
  relabelings.
  \end{minipage}
\end{table}

The high sparsity, symmetry and exact dyadic structure should be exploitable in nonlocality tests. One route is inflation techniques as in Ref.~\cite{Gitton2024}; another is symmetry-aware numerical search. We leave these explorations for future work.